# Portfolio optimization with idiosyncratic and systemic risks for financial networks

*This version: Nov 2021, any comments are welcomed*

Yajie Yang[1], Longfeng Zhao*[1], Lin Chen[1], Chao Wang[2], Jihui Han[3]

1. School of Management, Northwestern Polytechnical University, Xi'an, China
2. Research Base of Beijing Modern Manufacturing Development, College of Economics and Management, Beijing University of Technology, Beijing, China
3. College of Computer and Communication Engineering, Zhengzhou University of Light Industry, Zhengzhou, China

**Abstract**

In this study, we propose a new multi-objective portfolio optimization with idiosyncratic and systemic risks for financial networks. The two risks are measured by the idiosyncratic variance and the network clustering coefficient derived from the asset correlation networks, respectively. We construct three types of financial networks in which nodes indicate assets and edges are based on three correlation measures. Starting from the multi-objective model, we formulate and solve the asset allocation problem. We find that the optimal portfolios obtained through the multi-objective with networked approach have a significant over-performance in terms of return measures in an out-of-sample framework. This is further supported by the less drawdown during the periods of the stock market fluctuating downward. According to analyzing different datasets, we also show that improvements made to portfolio strategies are robust.

**Keyword** Portfolio optimization · Financial Networks ·Multi-objective· Dependence structures

**1 Introduction**

The mean-variance(MV) model proposed by Markowitz (1952) [1] laid the foundation of modern portfolio theory. As a static framework, the MV model revealed how to balance the mean return and the risk through portfolio diversification from the perspective of convex optimization. Since then, a vast amount of literature has focused on developing new concepts and techniques. Some extensions add various constraints to the MV problem, such as trading constraints (See Borko¹vec et al. (2010)[2] ; Chen et al. (2010)[3]; Brandes et al. (2012)[4]). Others contribute to provide good estimates for the expected returns and covariance matrix (See Athayde et al. (2002, 2004)[5][6] and Harvey et al. (2010)[7]). A more complete summary is provided by Kolm et al. (2014)[8], which reviewed the development and challenges and further research direction of mean-variance model. In addition to variance, the most widely used risk measure for extremely unfavorable outcomes is Vale-at-Risk (VaR), first provided by Morgan (1996)[9] . However, it has several undesirable theoretical properties (non-subadditivity and non-convexity). This led to the introduction of Conditional Value-at-Risk (CVaR). Since different risk measures quantify risk from different perspective, researchers have considered controlling multiple risk measures in portfolio selection process, such as mean-absolute deviation-skewness, mean-variance-skewness, mean-variance-VaR and mean-Variance-CVaR (See Konno et al. (1993)[10]; Konno et al. (1995) [11]; Wang (2000) [12]; Roman et al. (2007)[13]) These multi-objective models enable investors to balance risk and return under complex constraints.

[1] *Corresponding author
zlf@nwpu.edu.cn (Longfeng Zhao); zlfccnu@mails.ccnu.edu.cn(Longfeng Zhao)

It is well acknowledged that the financial market is a typical complex system. The financial assets' prices are formed according to the complicated interaction among investors' behavior, financial price fluctuations, financial contagion and many other factors. The classical Markowitz model, undoubtedly, can not fully quantify the influence of interconnection among assets by using the simple covariance matrix. Thanks to the powerful abstraction capability of the complex network, the interaction among assets can be directly mapped, with plenty correlation metrics, to network models (See Barigozzi et al. (2014)[14]; Billio et al. (2012)[15]; Bonanno et al. (2004)[16]; Diebold et al. (2014)[17]; Hautsch et al. (2015)[18]; Baitinger et al. (2017)[19]). Therefore, we characterize financial markets as networks in which individual assets are identified by nodes and the links among them portray the dependencies of returns. As such, correlation network is the most commonly used method to establish links among numerous asset classes and to specify which assets are important in terms of interconnectedness (based on some centrality measures). A thorough summary has been delivered by Marti et al. (2021)[20] . They review the correlations, hierarchies, networks and clustering in financial markets, regarding the financial network construction methodologies for the recent decades. The story began with Mantegna (1999)[22] who first proposed the concept of minimum spanning trees (MST) based on linear correlations between stocks to analyze the hierarchical structure in financial markets. Subsequently, Tumminello et al. (2005)[23] introduced the planar maximally filtered graph (PMFG) method and found that PMFG capture more information than MST. Since then, the correlation-based network or financial network has become the state-of-the-art to characterize the mutual dependency of financial instruments.

With the help of large strand of topological parameters from network theory, the correlations of asset returns as a risk measure has been increasingly recognized and the application of network theory in portfolio selection problems has attracted substantial attention. Onnela et al. (2003)[24] first studied the asset allocation problem under a network perspective, showing that the assets in the optimal Markowitz portfolio are always located on the outer leaves of the asset tree. Similarly, using MST and PMFG, Pozzi et al. (2013) [25] tested the portfolios by incorporating the financial network structural information. They proved that the over-weighting of peripheral stocks introduces higher Sharp ratios. For theoretical advancement, Peralta and Zareei (2016)[26] first established a link between Markowitz framework and the financial network, proving that optimal weights are negatively correlated with the centrality of assets, and used the centrality measures of constructed networks to make portfolio selection strategies. Výrost et al. (2019)[27] proposed further adjustments to portfolio selection strategies that utilize centrality measures from financial networks. Li et al. (2019)[28] studied the features of the network generated by the full cross-correlation matrix and the global-motion one, and their performances in the portfolio optimization. The previous mentioned researches on optimal allocation based on the correlation-based networks only optimized the asset selection procedure based on centrality measures, then make profitable investment decisions through them. Clemente et al. (2021)[29] first provided an objective function that incorporates the interconnectedness of returns into the portfolio optimization model. They were able to further improve the traditional Markowiz framework by defining a systemic connectedness metric based on the clustering coefficient of the fully connected correlation network and substitute the covariance matrix with the dependency matrix.

Through literature review, we notice that: (1) Mean-variance model is essentially a static framework, which neglects the dynamic changes of the correlations of returns and the adjustment of stock weights by investors. While in reality, investment behavior is a dynamic adjustment process. The contradiction between static model and dynamic investment leads to the necessity of dynamic portfolio strategy research. (2) A vast literature constructs the financial networks based on the linear relationship between two assets, then investigate the spreading of risk across the financial markets, the hierarchical structures of the stock exchange and the single-objective portfolio optimization. Although Clemente et al. (2021) employed the topological indicator into the objective function of the optimization model, the information of the volatility (variance) has been ignored. The direct substitution of the covariance matrix with the clustering coefficient matrix will inevitably undermine the portfolio riks. Thus, a network-based multi-objective optimization model deserves further attention.

Due to various of risk measures characterizing different risks, and the network theory effectively describing the complex interactions in the finance[30], in this work, we study the multi-objective portfolio optimization under a network perspective. We first construct the dependency networks by using Pearson correlation, Kendall's tau coefficient and Lower tail dependence. Then, by extending the work of Clemente et al. (2021)[29], we combine the clustering coefficient and variance-covariance matrix, corresponding to the dependency networks, leading to the mean-variance-systemic risk multi-objective optimal problem. By adjusting the trade-off parameter of our model, investors can flexibly balance the volatility of portfolio and systemic risk. Finally, we use the rolling window method to simulate the dynamic investment process, and employ the expected return, the *Omega ratio*, the proxy of transaction costs and the modified Herfindahl index to evaluate the portfolio performances. We advanced the growing literature on network applications in portfolio optimization problems (e.g., Onnela et al. (2003); Kaya (2015); Baitinger et al. (2016); López de Prado (2016); Peralta et al. (2016); Huang et al. (2018); Lwin et al. (2017)) by proposing the multi-objective portfolio optimization model, which combines the idiosyncratic and systemic risks characterized by variance and products of transitivity of assets respectively.

The following of the paper is organized as follows. Section 2 is concerned with the portfolio selection problem. Precisely, we explain three dependency structures used to construct networks and present multi-objective portfolio optimization with networked idiosyncratic and systemic risks. Section 3 describes the two datasets and empirical procedures. Section 4 performs in-sample and out-of-sample analysis for both our strategies and the benchmarks. Finally, Section 5 concludes and outlines future research directions.

## 2 Portfolio selection problems

### 2.1 Dependency network

In order to quantify the mutual dependency of assets, we sketch the stock time series through a complete, weighted, and undirected network $G=(V,E)$, where stocks are nodes and edges accounts for the dependence between returns. A network $G$ consists by a set $V$ of nodes and a set $E$ of links (edges). The edge connecting a pair of nodes $i$ and $j$ is denoted by $e_{ij}$. A network is weighted if a weight $w_{ij}$ is associated to each edge $e_{ij}$. In this case, weights $w_{ij} \in \mathbb{R}$ on the edges are described by the correlations of returns.

In our study, the linear dependence (Pearson correlation), and the non-linear dependence (including Kendall's tau coefficient and lower tail dependence) defined as follows are employed to capture the linkage among assets.

(1) The edge weight of the Pearson network is characterized by the Pearson correlation coefficient between stocks $i$ and $j$:

$$w_{ij} = \begin{cases} \rho_{ij}, & if\ i \neq j, \\ 0, & otherwise. \end{cases} \quad (1)$$

The Pearson coefficient between each couple of assets is computed as:

$$\rho_{ij} = \frac{\langle R_i R_j \rangle - \langle R_i \rangle \langle R_j \rangle}{\sigma_i \sigma_j}, \quad (2)$$

where $\langle \cdot \rangle$ is the expected value, $\sigma$ is the standard deviation.

(2) The edge weight of the Kendall Tau network is characterized by the Kendall's tau coefficient between stocks $i$ and $j$:

$$w_{ij} = \begin{cases} \omega_{ij}, & if\ i \neq j, \\ 0, & otherwise. \end{cases} \quad (3)$$

The Kendall' tau coefficient is computed as:

$$\tau_{ij} = \frac{\sum_{h=1}^{n} \sum_{k \neq h} \text{sgn}(R_{i,h} - R_{i,k}) \text{sgn}(R_{j,h} - R_{j,k})}{n(n-1)}, \quad (4)$$

where $R_{i,h}$ is the daily log return of the asset $i$ on the $h$ th trading day.

(3) The edge weight of the Tail dependency network is characterized by conditional probability of the lower tail distributions between stocks $i$ and $j$:

$$w_{ij} = \begin{cases} \eta_{ij}, & if\ i \neq j, \\ 0, & otherwise. \end{cases} \quad (5)$$

The lower tail dependence is defined as:

$$\eta_{ij} = \lim_{q \to 0} p(R_j \leq F_{R_j}^{-1}(q)\ |\ R_i \leq F_{R_i}^{-1}(q)), \quad (6)$$

where $F_{R_i}^{-1}(q)$ is the inverse of the cumulative distribution function of random variable $R_i$. Three different methods are introduced in order to extract the dependence structures among assets in a network context.

With the three mutual dependency metrics above, we can calculate the dependency matrices. Then, the dependency is used as the adjacency matrices of the corresponding dependency networks. It is straightforward that the dependency networks are weighted fully connected networks with undirected edges. The following portfolio optimization problems are based on these three dependency networks.

**2.2 Portfolio optimization through a multi-objective approach**

Following Markowitz (1952), the portfolio selection (mean-variance) can be defined as follows:

$$\text{P}: \begin{cases} \min\ \mathbf{x}^T \Sigma \mathbf{x} \\ \text{s.t}\ \ \mathbf{e}^T \mathbf{x} = 1 \\ \quad\ \ \mathbf{r}^T \mathbf{x} = r_p \\ \quad\ \ 0 \leq x_i \leq 1, i = 1, ..., N. \end{cases} \quad (7)$$

The formula expresses that we minimize the variance-covariance risk, where the matrix $\Sigma$ is an estimate of the variance and covariance of the assets. The vector $\mathbf{x}$ denotes the individual investments subject to the condition

$\mathbf{e}^T\mathbf{x} = 1$ ( $\mathbf{e}$ is a vector with all elements equal to 1) that available capital is fully invested (i.e., budget constraint). The expected return $r_p$ is expressed by the condition $\mathbf{r}^T\mathbf{x} = r_p$.

Clemente et al. (2021) proposed a network-based asset allocation through defining the N-square matrix $C$, interpreted as an interconnectedness matrix, whose elements are:

$$c_{ij} = \begin{cases} C_i C_j, & \text{if } i \neq j \\ 1, & \text{otherwise,} \end{cases} \quad (8)$$

where $C_i$ is the average clustering coefficient for a node $i$. Thus, matrix $C$ accounts for the interconnection level of each asset with the system, instead of measuring classical pairwise correlation. Then substitute the covariance matrix of the assets with the network-based dependency matrix, leading to the objective function containing the topology information. As a result, the modified model neglect the information delivered by the variance-covariance matrix of the system..

In view of this, we further propose the multi-objective model with two risk measures, the idiosyncratic (variance) and systemic (the clustering coefficient) risks. To assess the investment performance of our models, in this section, we add the median of the mean return as a constraint to Clemente et al.'s model (here we refer as CGP) and MV. Furthermore, the MV and CGP are used as two comparison models.

We assume that $N$ assets are available. And let $\mathbf{r} = (r_1, r_2, ..., r_N)^T$ be the daily log return vector, $\mathbf{x} = (x_1, x_2, ..., x_N)^T$ the vector of the portfolio weights. The portfolio optimization problem can be defined as follow:

$$P: \begin{cases} \min \left( \alpha \mathbf{x}^T H \mathbf{x} + (1-\alpha) \mathbf{x}^T \Sigma \mathbf{x} \right) \\ \text{s.t } \mathbf{e}^T \mathbf{x} = 1 \\ \mathbf{r}^T \mathbf{x} = r_p \\ 0 \leq x_i \leq 1, i = 1, ..., N. \end{cases} \quad (9)$$

Here, a model $P$ for multi-objective portfolio selection is proposed by incorporating both the clustering coefficient matrix and the variance-covariance matrix. We claim that taking two risks into consideration, instead of one, gives a complete description of the risk. Here $H = \Delta^T C \Delta$, where $\Delta$ is a diagonal matrix with diagonal entries $\sigma$, the standard deviation of the asset return. The interconnectedness matrix $C$ with the principal diagonal entry is constant 1 and others are the product of the average clustering coefficient for each couple assets as given in Eq(8), which describe the coupling strengthen of each asset with the system. In particular, $C$ being dependent on a network-based measure of systemic risk (i.e., the clustering coefficient) results in $H$ implicitly including a measure of the state of stress of the financial system. The measure of idiosyncratic risk is generated by the covariance matrix of the returns of all the assets $\Sigma$, in which the entries are referred to as idiosyncratic volatility of the assets. A constant $\alpha$ ( $\alpha \in [0,1]$ ), represents the attitude of investors towards the variance of the portfolio and the systemic interconnectedness level. The larger the threshold $\alpha$ is, the more sensitive the investor is to the systemic risk. Conditions $0 \leq x_i \leq 1$ exclude the possibility of short selling, which is inline with the Chinese market situation.

## 3 Data and methodology

### 3.1 Data description

In this section, we perform some empirical applications to assess the effectiveness of the proposed approaches. For the sake of robustness, we analyzed two portfolios arising from two datasets，the constitute stocks for both SSE 180 index and SSE 50 index. Since the SSE 180 index is composed of 180 stocks with large scale, good liquidity, and industry representativeness. We collect daily returns of the dataset that includes 126 leading Chinese stocks constituents of the SSE 180 index in the time-period ranging from April 2014 to August 2021. In addition, we collect daily returns of a second dataset for the same time period that includes 42 stocks constituents of the SSE 50 index. Both datasets are downloaded from Joint Quant.

**3.2 Empirical procedure**

In the literature, the performance of a portfolio can be evaluated from diverse aspects. In this article, we focus on portfolio diversification, transaction cost and return performance measures, as will be explained in the following subsections. All these measurements are investigated through a rolling window methodology. We construct windows for 12 months that are rolled one month ahead, resulting $w = 1, 2, ..., M$ overlapping windows.

The empirical procedure will be processed according to the following procedure.

Step 1: determine the width $n$ of the in-sample window and the width $h$ of the out-of-sample period. The data of the first in-sample window (i.e., from $t = 1$ to $t = n$) are used to perform a Pearson Correlation network approach (PNA), Kendall tau network approach and Tail dependence network approach (TNA) and compute the corresponding matrix $H$.

Step 2: let the value of $r_p$ be the median of the mean returns of the assets in the current window. And consider twenty-one levels of $\alpha$, referred to as dividing the interval $[0,1]$ into 20 equal parts. It is worth noting that, when $\alpha$ equals 0 and 1, the optimal problem $P$ is transformed into the mean-variance model (MV) and the portfolio selection problem CGP (i.e., the median of the mean return constraint is further considered on the basis of Clemente et al. (2021)), respectively. A natural choice is to compare our results with the strategies of MV and CGP. To this reason, we use them as two benchmarks.

Step 3: based on the given $r_p$ and $\alpha$, the optimal problem can be solved by the cvx toolkit, and the optimal weights obtained are then invested in the out-of-sample period.

Step 4: the process is repeated for the next rolling window $h$ steps forward, until the end of the dataset is reached. We buy-and-hold the portfolios and record the in-sample and out-of-sample performances in each rebalancing period.

**3.3 Performance measures**

We analyze the obtained strategies based on several portfolio-specific measures: the modified Herfindahl index ($HI_w$), the turnover ($\vartheta$), the break-even transaction cost (BETC), the drawdown ($D$), the expected value ($\varphi(w)$) and the Omega ratio (*OR*).

The diversification of each portfolio is evaluated by the modified Herfindahl index ($HI_w$). Given strategy $\mathbf{x}^*$, standing for the optimal weights on window $w$, $HI_w$ is defined as:

$$HI_w = \frac{\mathbf{x}^{*T}\mathbf{x}^* - 1/N}{1 - 1/N}, \tag{10}$$

a smaller $HI_w$ indicates a more diversified portfolio. In particular, being equal to 0, in case of the equally weighted portfolio (EW, the most diversified portfolio) and to 1 in case of a portfolio concentrated in only one asset.

One proxy of the transaction costs of a given strategy is the portfolio turnover ($\vartheta$), which is computed as:

$$\vartheta(w) = \sum_{i=1}^{N} |x_i^* - x_i^-|, \tag{11}$$

where $x_i^-$ and $x^*$ are the optimal portfolio weights of the $i-th$ asset before and after rebalancing (i.e., according to the optimization strategy) at window $w$, respectively.

It is also important to investigate the impact of transaction costs. Following the approach provided in Han et al. (2013)[22], we compute the breakeven transaction cost (BETC). The BETC is computed as follows:

$$BETC(w) = h\mathbf{e}^T(R_{out} * \mathbf{x}^*), \tag{12}$$

where $R_{out}$ is the out-of-sample return on window $w$.

If $P_{i,w}$ is the value of the $i-th$ portfolio at $w$ time, the drawdown ($D$) is defined as:

$$D_{i,w} = \frac{P_{i,w}}{\max_{t=1,..,w} P_{i,t}}. \tag{13}$$

In this paper, we examine the return performances of the two datasets by using the expected value and Omega ratio (*OR*). The expected value ($\varphi(w)$) is defined as:

$$\varphi(w) = E[R_{out} * \mathbf{x}^*]. \tag{14}$$

Compared with the Sharp Ratio, which only includes the first and second moments of the return distribution, the Omega ratio makes full use of the distribution information, thus, it can better evaluate the performance of the investment strategy (See Keating et al. (2002)[36]). It is defined as:

$$OR = \frac{\int_\varepsilon^\infty (1-F(x))dx}{\int_{-\infty}^\varepsilon F(x)dx} = \frac{E(R_{out} * \mathbf{x}^*)}{E(\varepsilon - R_{out} * \mathbf{x}^*)}, \tag{15}$$

where $F(x)$ is the cumulative distribution function of the portfolio returns and $\varepsilon$ is a specified threshold. Returns below the specific threshold are considered as losses and returns above as gains. For the sake of simplicity, we set $\varepsilon$ at zero.

## 4 Results and discussion

In this paper, we focus on monthly stepped 1-year window and perform rolling network-based strategies by considering 75 sub-samples (i.e., 75 rolling windows) of the initial dataset. By means of the optimal weights obtained by each approach, we can compute the portfolio diversification, the proxy of transaction costs and the drawdown. Moreover, by investing the optimal weights in the out-of-sample periods, we also provide return measures.

### 4.1 SSE 180 dataset with monthly stepped 1-year rolling windows

Regarding SSE 180, we calculate the average of performance measures described in Sect. 3.2 over $w = 1,2,...,M$ respectively, main results reported in Table 1. We observe PNA in Table 1 that, in general, *P* portfolios led to better expected return and *Omega ratio* (*OR*) than MV and CGP. Meanwhile, the turnover is larger than MV approach, which means that allocated weights have been rebalanced more than MV. One might question whether the improved return portfolio characteristics are offset by increased transaction costs. In fact, it is well acknowledged that the cost is larger for most individual network-based asset allocation strategies (See Peralta et al. (2016)[26], Vyrost et al. (2019)[27][13], DeMiguel et al. (2009)[38]). Furthermore, on this specific dataset, it is interesting to note that the optimal network-based portfolios are increasingly diversified with $\alpha$ decreasing. This empirical evidence seems reasonable, since relaxing constraint on systemic risk means a low level of clustering which leads to more diversified portfolios. By observing KNA and TNA in Table 1, we get similar results.

**Table 1** SSE 180 portfolio results for different settings of $\alpha$, with 1 year in-sample and 1 month out-of-sample. We report the out-of-sample expected value ($\varphi$), Omega ratio ($OR$), and in-sample turnover ($\vartheta$), modified Herfindahl index ($HI$) for portfolio strategies.

| | | | | | PNA | | | | | | |
|---|---|---|---|---|---|---|---|---|---|---|---|
| $\alpha$ | 0 | 0.05 | 0.1 | 0.15 | 0.2 | 0.25 | 0.3 | 0.35 | 0.4 | 0.45 | 0.5 |
| $\varphi$ | -0.01668 | -0.00553 | -0.00522 | -0.00490 | -0.00458 | -0.00425 | -0.00395 | -0.00360 | -0.00323 | -0.00266 | -0.00222 |
| $OR$ | 0.95446 | 0.98495 | 0.98582 | 0.98668 | 0.98755 | 0.98846 | 0.98927 | 0.99024 | 0.99125 | 0.99279 | 0.99400 |
| $\vartheta$ | 0.43554 | 0.48560 | 0.48758 | 0.48951 | 0.49159 | 0.49361 | 0.49562 | 0.49801 | 0.50099 | 0.50358 | 0.50627 |
| $HI$ | 0.14588 | 0.20350 | 0.20475 | 0.20604 | 0.20739 | 0.20880 | 0.21025 | 0.21172 | 0.21327 | 0.21484 | 0.21648 |
| $\alpha$ | 0.55 | 0.6 | 0.65 | 0.7 | 0.75 | 0.8 | 0.85 | 0.9 | 0.95 | 1 | |
| $\varphi$ | -0.00185 | -0.00157 | -0.00137 | -0.00126 | -0.00134 | -0.00159 | -0.00182 | -0.00188 | -0.00196 | -0.00183 | |
| $OR$ | 0.99501 | 0.99575 | 0.99631 | 0.99662 | 0.99640 | 0.99576 | 0.99518 | 0.99505 | 0.99488 | 0.99524 | |
| $\vartheta$ | 0.50926 | 0.51171 | 0.51407 | 0.51662 | 0.52002 | 0.52438 | 0.52978 | 0.53601 | 0.54269 | 0.55112 | |
| $HI$ | 0.21648 | 0.21828 | 0.22018 | 0.22206 | 0.22409 | 0.22619 | 0.22857 | 0.23136 | 0.23440 | 0.23763 | |
| | | | | | KNA | | | | | | |
| $\alpha$ | 0 | 0.05 | 0.1 | 0.15 | 0.2 | 0.25 | 0.3 | 0.35 | 0.4 | 0.45 | 0.5 |
| $\varphi$ | -0.00211 | 0.00059 | 0.00065 | 0.00064 | 0.00058 | 0.00049 | 0.00043 | 0.00040 | 0.00037 | 0.00040 | 0.00046 |
| $OR$ | 0.99417 | 1.00162 | 1.00178 | 1.00175 | 1.00159 | 1.00135 | 1.00118 | 1.00110 | 1.00102 | 1.00110 | 1.00126 |
| $\vartheta$ | 0.41860 | 0.46088 | 0.46190 | 0.46269 | 0.46367 | 0.46479 | 0.46597 | 0.46736 | 0.46901 | 0.47075 | 0.47275 |
| $HI$ | 0.10474 | 0.15964 | 0.16088 | 0.16216 | 0.16350 | 0.16491 | 0.16636 | 0.16790 | 0.16948 | 0.17113 | 0.17282 |
| $\alpha$ | 0.55 | 0.6 | 0.65 | 0.7 | 0.75 | 0.8 | 0.85 | 0.9 | 0.95 | 1 | |
| $\varphi$ | 0.00062 | 0.00080 | 0.00097 | 0.00100 | 0.00095 | 0.00092 | 0.00075 | 0.00040 | 0.00009 | -0.00006 | |
| $OR$ | 1.00169 | 1.00216 | 1.00264 | 1.00270 | 1.00257 | 1.00246 | 1.00202 | 1.00106 | 1.00023 | 0.99984 | |
| $\vartheta$ | 0.47480 | 0.47709 | 0.47993 | 0.48288 | 0.48601 | 0.49069 | 0.49623 | 0.50250 | 0.50985 | 0.51776 | |
| $HI$ | 0.17457 | 0.17637 | 0.17831 | 0.18036 | 0.18244 | 0.18451 | 0.18674 | 0.18902 | 0.19155 | 0.19436 | |
| | | | | | TNA | | | | | | |
| $\alpha$ | 0 | 0.05 | 0.1 | 0.15 | 0.2 | 0.25 | 0.3 | 0.35 | 0.4 | 0.45 | 0.5 |
| $\varphi$ | -0.00411 | -0.00026 | -0.00024 | -0.00027 | -0.00031 | -0.00032 | -0.00035 | -0.00025 | -0.00019 | -0.00022 | -0.00019 |
| $OR$ | 0.98909 | 0.99929 | 0.99935 | 0.99925 | 0.99916 | 0.99912 | 0.99906 | 0.99932 | 0.99949 | 0.99939 | 0.99950 |
| $\vartheta$ | 0.46436 | 0.48710 | 0.48878 | 0.49050 | 0.49232 | 0.49437 | 0.49649 | 0.49921 | 0.50245 | 0.50600 | 0.50987 |
| $HI$ | 0.11157 | 0.18795 | 0.18977 | 0.19161 | 0.19354 | 0.19559 | 0.19773 | 0.19990 | 0.20212 | 0.20449 | 0.20702 |
| $\alpha$ | 0.55 | 0.6 | 0.65 | 0.7 | 0.75 | 0.8 | 0.85 | 0.9 | 0.95 | 1 | |
| $\varphi$ | -0.00017 | -0.00020 | -0.00016 | 0.00005 | 0.00017 | 0.00019 | -0.00008 | -0.00054 | -0.00107 | -0.00152 | |
| $OR$ | 0.99956 | 0.99945 | 0.99958 | 1.00013 | 1.00045 | 1.00050 | 0.99979 | 0.99858 | 0.99723 | 0.99610 | |
| $\vartheta$ | 0.51367 | 0.51736 | 0.52086 | 0.52459 | 0.52835 | 0.53302 | 0.53878 | 0.54590 | 0.55278 | 0.56146 | |
| $HI$ | 0.20966 | 0.21246 | 0.21540 | 0.21853 | 0.22182 | 0.22537 | 0.22927 | 0.23368 | 0.23871 | 0.24434 | |

According to Table 1, the best portfolio cannot be clearly identified. Since $OR$ implicitly embodies all moments of return distribution without any priori assumption, we define the dominance ratio that the fraction of the difference between the Omega ratio of our model $P$ and CGP model (i.e., $OR = (OR^P - OR^{CGP})/OR^{CGP}$). For illustrative purposes, we further concern heatmaps (a), (b) and (c) from Fig.1, plotting the out-of-sample dominance ratio. The larger the value, the better the performance of our model. Heatmap from Fig.1(a) presenting a scenario PNA, in which for each window we can always find an optimal $\alpha$ leading to better out-of-sample performance than CGP model in terms of *OR* (i.e., the redder the color). It is suggested that our model overperforms CGP model. Then, we provide a new optimal investment strategy $P^*$ by selecting the weights corresponding to the best performance $\alpha$ at each window. We show the performance measures of the $P^*$ strategy in Fig. 2 and Table 2, using MV and CGP model as two benchmarks.

In Fig. 2 (a)-(c), the cumulative returns for the investment strategy $P^*$, CGP and MV are displayed together with the return of the market index denoted SSE180 for reference. Through the comparison of the cumulative returns, we observe that the strategy $P^*$ based on our optimization scheme obtained a significant performance imporvement with respect to MV and CGP models under three network-based portfolios.

The Fig.2 (d)-(f) clearly show that the benchmark strategies MV and CGP led to the worst cumulative returns most of the time and suffered from excessive drawdown periods. Regarding to the portfolio performance measures described above, it seems that our asset allocation scheme $P^*$ might be valuable for investors to adapt to the market environment.

Meanwhile, in Fig.2 (g)-(i), our strategies possess larger turnover in some windows. Comparing with the first panel in Fig.2, clearly, $P^*$ will correspond to a higher turnover when the market index suffered from fluctuated downward periods. It seems reasonable that when the market switches its styles, we also need to readjust our strategies to capture new investment opportunities. This indicates that there is a trade-off between returns and turnover.

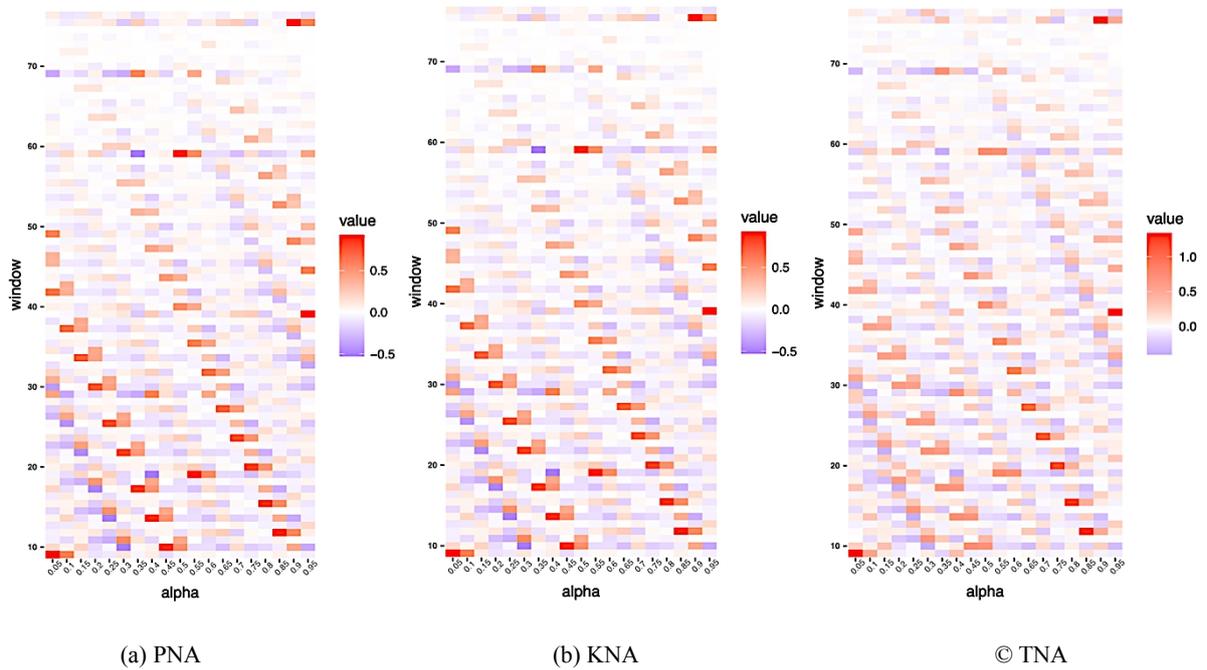

(a) PNA          (b) KNA          © TNA

**Fig.1** SSE 180 dominance ratio heatmaps. The $x$-axes represent the value of $\alpha$ and the y-axes describe each window (i.e., 75 rolling windows).

For the sake of completeness, we also provide the quantitative performance measures in Table 2. The results obtained from expected return and *Omega ratio* indicate that the over-performance of three dependence network approaches with respect to MV and CGP. In particular, it is noticeable the *OR* of $P^*$ that is higher than 1 in case of three dependence networks since the expected gains are higher than the expected loss, but the other two strategies just the opposite of $P^*$. This benefit is further supported by the average drawdown tending to be smaller, while the *OR* is always higher for $P^*$ strategies.

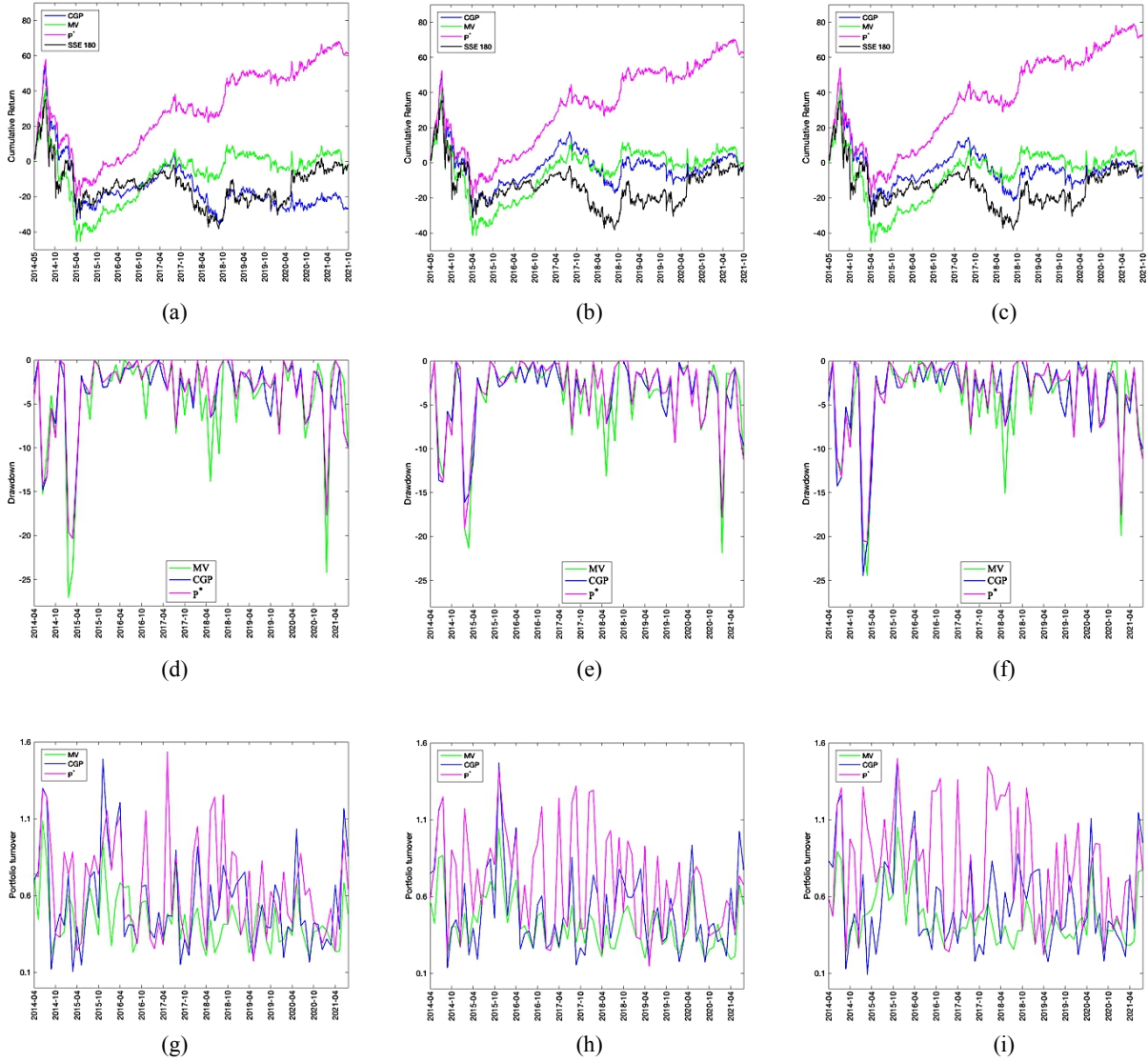

**Fig.2** Cumulative returns, drawdowns and turnovers of SSE 180 portfolio strategies. The dates on the $x$-axes represent the initial dates of the buy and hold strategy of out-of-sample windows $w$. (a)-(b) Cumulative returns for PNA, KNA and TNA respectively. (d)-(f) Drawdowns for PNA, KNA and TNA. The y-axes are percentages. (g)-(i) Turnover for PNA, KNA and TNA.

In Table 2, it is evident that the turnover of $P^*$ is slightly larger under three dependence networks. As noted above, the cost is larger for most network-based asset allocation strategies. Generally, a strategy with a high turnover rate exhibits higher costs. However, the increased turnover of $P^*$ seems to be more than compensated by increased returns. This statement supported by the break-even transaction cost (BETC), the values under PNA, KNA and TNA are 82.59bps (basis point), 84.64bps and 97.67bps, respectively. As stated in Balduzzi and Lynch[24], reasonable lower and upper bounds for the transition cost are 1bp and 50bps, respectively. Thus, obtain a BETC higher than 50bps means that a disproportionally abnormal transaction cost is required to wipe out the returns of the $P^*$ strategy. That is to say, we wish the BETC of the ideal portfolio to be greater than 50bps. The larger the BETC, the more difficult it is for the improved returns of $P^*$ to be offset by increased transaction costs (See Wang et al. (2021)[35], Vyrost et al. (2019)[27], Peralta et al. (2016)[26]). Therefore, our calculations indicate large values of BETC which supports that the outperformance of network-based multi-objective portfolios $P^*$ is robust with respect to transaction costs.

Furthermore, as a consequence of Table 2, the optimal strategy $P^*$ is more diversified than CGP. Specially referring to systemic risk, financial stability, and contagion, it is of importance to diversify risk.

**Table 2** Performance of portfolio strategies. Report performance measurements for the $P^*$ strategy consisting of the weights corresponding to the best performance $\alpha$ at each window.

| strategy | PNA | | | KNA | | | TNA | | |
|---|---|---|---|---|---|---|---|---|---|
| | $P^*$ | MV | CGP | $P^*$ | MV | CGP | $P^*$ | MC | CGP |
| $\varphi$ | 0.0393 | -0.0167 | -0.0018 | 0.0403 | -0.0021 | -0.0000 | 0.0465 | -0.0041 | -0.0015 |
| OR | 1.1121 | 0.9545 | 0.9952 | 1.1155 | 0.9942 | 0.9998 | 1.1311 | 0.9891 | 0.9961 |
| D | -0.0349 | -0.0423 | -0.0382 | -0.0344 | -0.0399 | -0.0373 | -0.0368 | -0.0415 | -0.0401 |
| $\vartheta$ | 0.6524 | 0.4355 | 0.5511 | 0.6971 | 0.4186 | 0.5178 | 0.8031 | 0.4644 | 0.5615 |
| HI | 0.2000 | 0.1459 | 0.2411 | 0.1492 | 0.1047 | 0.1944 | 0.1717 | 0.1116 | 0.2443 |

## 4.2 SSE 50 dataset with monthly stepped 1-year rolling windows

The different scale of the stock market, SSE 50, has also been analyzed. We report main results obtained by using monthly stepped 1-year window in Table 3.

Regarding the performances, one interesting finding is that, on the whole, the return performances of $P$ strategies are gradually decreasing with the increase of $\alpha$. A possible explanation for this might be that the stock market is positive, compared to systemic risk, investors tend to pay more attention to the idiosyncratic volatility of the assets, because of high stakes, high returns. Thus, MV strategies provide the best behavior for PNA and TNA. We also note that the different kinds of dependence structures do not affect the performance in a significant way. For example, all the considered $\alpha$ s generate better returns for TNA. The multi-objective portfolios under a network perspective $P$, in the majority of cases, are characterized by much higher returns with respect to the CGP portfolio.

**Table 3** SSE 50 portfolio results for different settings of $\alpha$, with 1 year in-sample and 1 month out-of-sample. We report the out-of-sample expected value ($\varphi$), Omega ratio (OR), and in-sample turnover ($\vartheta$), modified Herfindahl index (HI) for portfolio strategies.

| | PNA | | | | | | | | | | |
|---|---|---|---|---|---|---|---|---|---|---|---|
| $\alpha$ | 0 | 0.05 | 0.1 | 0.15 | 0.2 | 0.25 | 0.3 | 0.35 | 0.4 | 0.45 | 0.5 |
| $\varphi$ | 0.01999 | 0.01600 | 0.01579 | 0.01554 | 0.01533 | 0.01515 | 0.01508 | 0.01498 | 0.01489 | 0.01477 | 0.01465 |
| OR | 1.05061 | 1.03973 | 1.03915 | 1.03849 | 1.03792 | 1.03744 | 1.03724 | 1.03696 | 1.03671 | 1.03638 | 1.03604 |
| $\vartheta$ | 0.37515 | 0.40992 | 0.41079 | 0.41170 | 0.41272 | 0.41387 | 0.41505 | 0.41629 | 0.41776 | 0.41948 | 0.42139 |
| HI | 0.10440 | 0.14282 | 0.14399 | 0.14522 | 0.14654 | 0.14793 | 0.14940 | 0.15096 | 0.15264 | 0.15440 | 0.15624 |
| $\alpha$ | 0.55 | 0.6 | 0.65 | 0.7 | 0.75 | 0.8 | 0.85 | 0.9 | 0.95 | 1 | |
| $\varphi$ | 0.01446 | 0.01423 | 0.01404 | 0.01387 | 0.01378 | 0.01380 | 0.01386 | 0.01388 | 0.01386 | 0.01386 | |
| OR | 1.03554 | 1.03493 | 1.03440 | 1.03396 | 1.03371 | 1.03376 | 1.03387 | 1.03387 | 1.03377 | 1.03370 | |
| $\vartheta$ | 0.42372 | 0.42590 | 0.42857 | 0.43167 | 0.43496 | 0.43843 | 0.44222 | 0.44734 | 0.45391 | 0.46147 | |
| HI | 0.15825 | 0.16037 | 0.16249 | 0.16479 | 0.16722 | 0.16974 | 0.17236 | 0.17523 | 0.17837 | 0.18166 | |
| | KNA | | | | | | | | | | |
| $\alpha$ | 0 | 0.05 | 0.1 | 0.15 | 0.2 | 0.25 | 0.3 | 0.35 | 0.4 | 0.45 | 0.5 |
| $\varphi$ | 0.01614 | 0.01834 | 0.01827 | 0.01819 | 0.01810 | 0.01804 | 0.01813 | 0.01825 | 0.01829 | 0.01833 | 0.01834 |

| | | | | | | | | | | |
|---|---|---|---|---|---|---|---|---|---|---|
| OR | 1.04022 | 1.04558 | 1.04538 | 1.04516 | 1.04492 | 1.04474 | 1.04496 | 1.04523 | 1.04529 | 1.04538 | 1.04536 |
| $\vartheta$ | 0.41860 | 0.46088 | 0.46190 | 0.46269 | 0.46367 | 0.46479 | 0.46597 | 0.46736 | 0.46901 | 0.47075 | 0.47275 |
| HI | 0.10474 | 0.15964 | 0.16088 | 0.16216 | 0.16350 | 0.16491 | 0.16636 | 0.16790 | 0.16948 | 0.17113 | 0.17282 |
| $\alpha$ | 0.55 | 0.6 | 0.65 | 0.7 | 0.75 | 0.8 | 0.85 | 0.9 | 0.95 | 1 | |
| $\varphi$ | 0.01829 | 0.01814 | 0.01796 | 0.01771 | 0.01754 | 0.01731 | 0.01714 | 0.01704 | 0.01716 | 0.01724 | |
| OR | 1.04519 | 1.04478 | 1.04429 | 1.04361 | 1.04313 | 1.04250 | 1.04204 | 1.04173 | 1.04194 | 1.04207 | |
| $\vartheta$ | 0.47480 | 0.47709 | 0.47993 | 0.48288 | 0.48601 | 0.49069 | 0.49623 | 0.50250 | 0.50985 | 0.51776 | |
| HI | 0.17457 | 0.17637 | 0.17831 | 0.18036 | 0.18244 | 0.18451 | 0.18674 | 0.18902 | 0.19155 | 0.19436 | |
| TNA | | | | | | | | | | | |
| $\alpha$ | 0 | 0.05 | 0.1 | 0.15 | 0.2 | 0.25 | 0.3 | 0.35 | 0.4 | 0.45 | 0.5 |
| $\varphi$ | 0.01975 | 0.01958 | 0.01960 | 0.01957 | 0.01952 | 0.01940 | 0.01928 | 0.01910 | 0.01902 | 0.01896 | 0.01885 |
| OR | 1.04840 | 1.04824 | 1.04828 | 1.04820 | 1.04805 | 1.04774 | 1.04746 | 1.04700 | 1.04677 | 1.04660 | 1.04631 |
| $\vartheta$ | 0.44716 | 0.47511 | 0.47653 | 0.47779 | 0.47907 | 0.48064 | 0.48199 | 0.48352 | 0.48542 | 0.48685 | 0.48813 |
| HI | 0.16237 | 0.20935 | 0.21034 | 0.21136 | 0.21248 | 0.21367 | 0.21490 | 0.21618 | 0.21755 | 0.21885 | 0.22014 |
| $\alpha$ | 0.55 | 0.6 | 0.65 | 0.7 | 0.75 | 0.8 | 0.85 | 0.9 | 0.95 | 1 | |
| $\varphi$ | 0.01871 | 0.01845 | 0.01813 | 0.01772 | 0.01720 | 0.01665 | 0.01637 | 0.01630 | 0.01608 | 0.01583 | |
| OR | 1.04595 | 1.04529 | 1.04447 | 1.04344 | 1.04211 | 1.04072 | 1.03998 | 1.03976 | 1.03915 | 1.03846 | |
| $\vartheta$ | 0.48981 | 0.49177 | 0.49403 | 0.49675 | 0.50020 | 0.50429 | 0.50802 | 0.51352 | 0.52216 | 0.53221 | |
| HI | 0.22156 | 0.22312 | 0.22479 | 0.22646 | 0.22822 | 0.23012 | 0.23226 | 0.23451 | 0.23721 | 0.23996 | |

We also plot the heatmaps Fig. 3, displaying the out-of-sample dominance ratio. Similarly, we construct the optimal investment strategy $P^*$. As expected, the cumulative out-of-sample returns of the $P^*$ portfolios outperform the MV and CGP portfolios (see Fig. 3 (a)-(b)). It is evident in Table 4 that both return measures of $P^*$ confirm the over-performance under three dependence structures, compared to MV and CGP. Meanwhile, the *Omega ratio* is higher than 1 because the expected gains are higher than the expected losses. The improved performances are also confirmed by the smaller drawdown of $P^*$. Overall, the asset allocation strategies in this dataset do lead to better return outcomes with respect to Table 2.

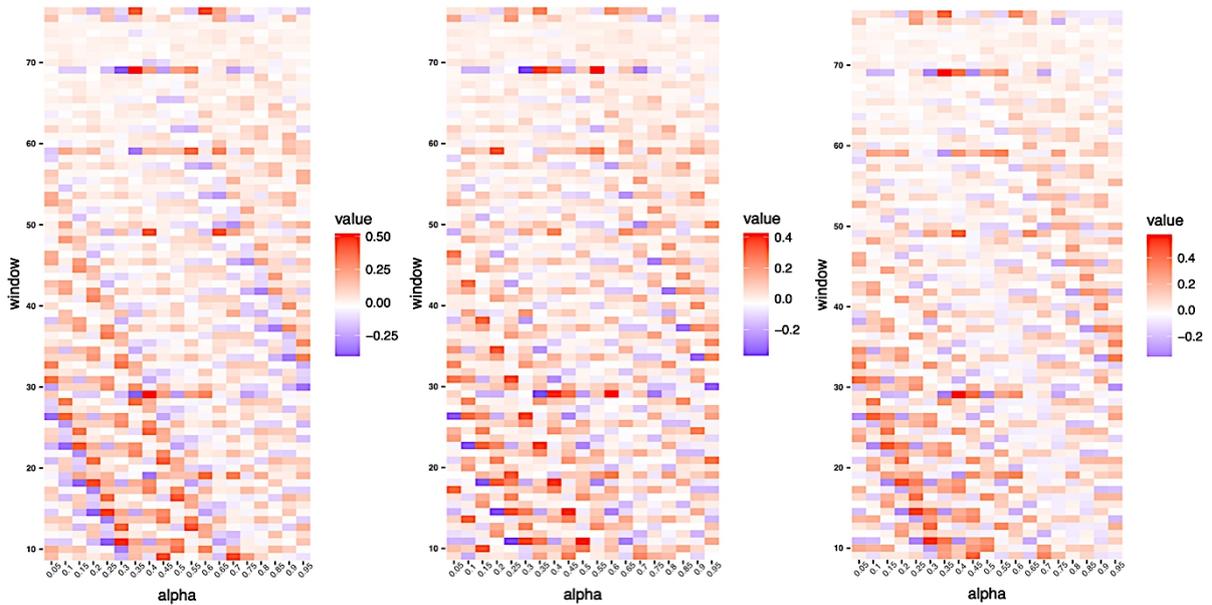

(a) PNA                    (b) KNA                  (c) TNA

**Fig.3** SSE 50 dominance ratio heatmaps. The $x$-axes represent the value of $\alpha$ and the y-axes describe each window (i.e., 75 rolling windows).

During the periods of the market index fluctuated downward, the benchmark strategies MV and CGP suffered from excess drawdown (See Fig.4 (d)-(f)). At the same time, our strategies $P^*$ manage to earn a higher level of turnover to readjust portfolio configurations to get with style rotation of stocks.

As already discussed, for most network-based portfolios, the cost is larger. As a consequence of Table 4, the turnover of $P^*$ is greater than CGP but worse than MV. Furthermore, we calculate the BETC of $P^*$ under PNA, KNA and TNA, corresponding to 135.48bps, 124.69bps and 139.61bps. All of them are higher than 50bps, leading to the improved return of $P^*$ is robust with respect to transaction costs. We also notice that the diversification of $P^*$ is higher than CGP. These findings obtained in case of the SSE 50 are in line with the previous dataset.

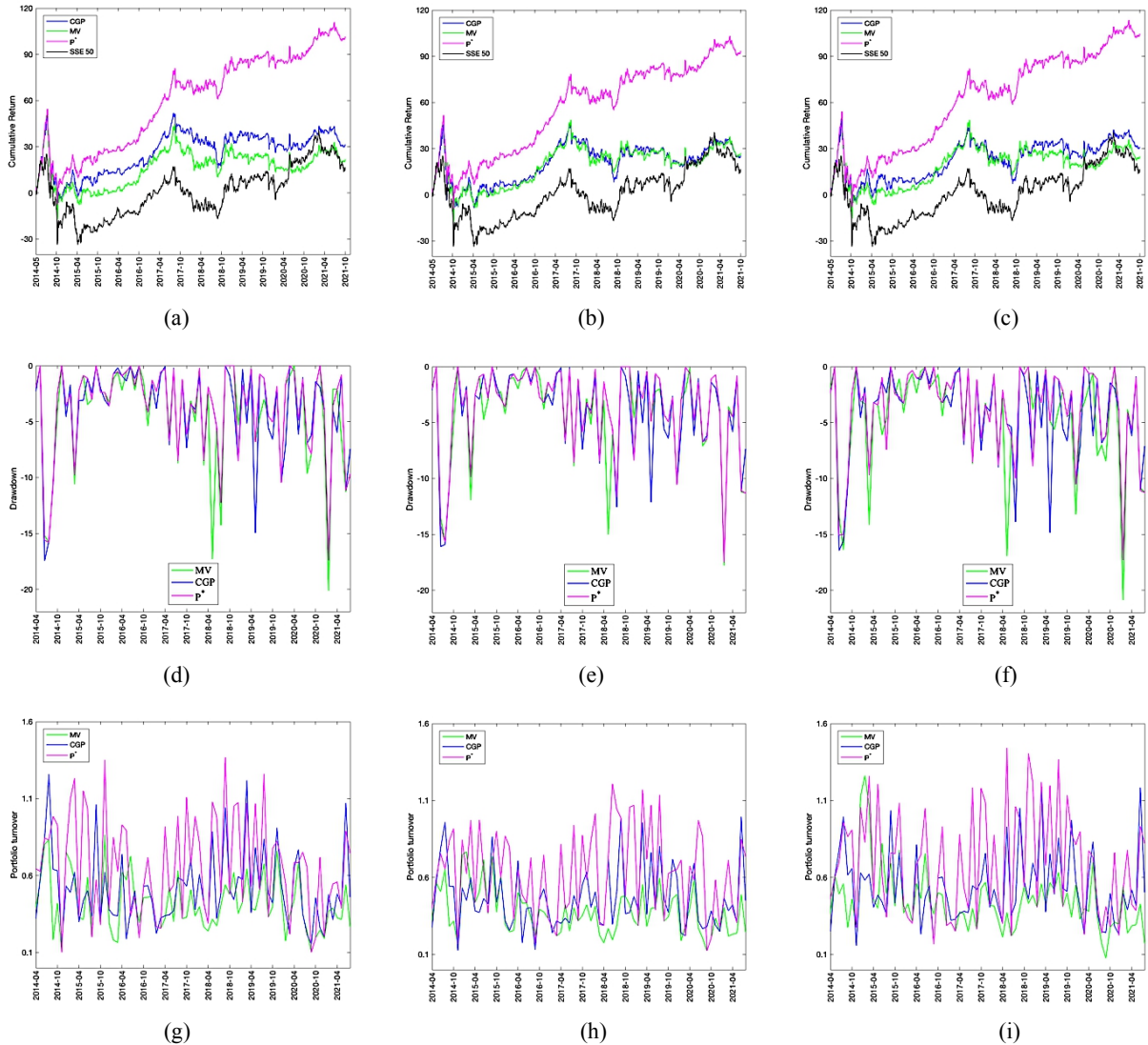

**Fig.4** Cumulative returns, drawdowns and turnovers of SSE 50 portfolio strategies. The dates on the $x$-axes represent the initial dates of the buy and hold strategy of out-of-sample windows $w$. (a)-(b) Cumulative returns for PNA, KNA and TNA respectively. (d)-(f) Drawdowns for PNA, KNA and TNA. The y-axes are percentages. (g)-(i) Turnover for PNA, KNA and TNA.

**Table 4** Performance of portfolio strategies. Report performance measurements for the $P^*$ strategy consisting of the weights corresponding to the best performance $\alpha$ at each window.

| strategy | PNA | | | KNA | | | TNA | | |
|---|---|---|---|---|---|---|---|---|---|
| | $P^*$ | MV | CGP | $P^*$ | MV | CGP | $P^*$ | MC | CGP |
| $\varphi$ | 0.0645 | 0.0200 | 0.0139 | 0.0594 | 0.0161 | 0.0172 | 0.0665 | 0.0198 | 0.0158 |
| OR | 1.1716 | 1.0506 | 1.0377 | 1.1446 | 1.0402 | 1.0421 | 1.1730 | 1.0484 | 1.0385 |
| D | -0.0365 | -0.0425 | -0.0408 | -0.0364 | -0.0409 | -0.0400 | -0.0373 | -0.0456 | -0.0414 |
| $\vartheta$ | 0.6725 | 0.4188 | 0.5093 | 0.6346 | 0.3752 | 0.4615 | 0.6978 | 0.4472 | 0.5322 |
| HI | 0.2055 | 0.1732 | 0.2420 | 0.1406 | 0.1044 | 0.1817 | 0.2012 | 0.1624 | 0.2400 |

Now, we summarize main insights observed on two datasets (SSE 180 and SSE 50) we have analyzed. Independently from the scale of dataset, the network-based multi-objective strategies have greater return performances and the larger diversification than the benchmark one (CGP). The turnover is higher for all network-based portfolios. However, the increased turnover seems to be compensated for by increase returns, since the BETC is larger for our strategies. We therefore conclude that our portfolios are robust to different scale of portfolio configurations.

## 5 Conclusion

In this paper, we propose a new multi-objective portfolio optimization with idiosyncratic and systemic risks for financial networks extension of standard Markowitz and CGP portfolio strategies. The objective function combines the traditional risk, referred to as the idiosyncratic volatility, and the systemic risk derived from the asset correlation networks. The two risk measures described by the variance (covariance) and clustering coefficient, respectively. According to adjusting the trade-off parameter $\alpha$, investors can balance the preference for volatility and systemic risk. Three types of dependence network approaches, Pearson Correlation network approach (PNA), Kendall tau network approach (KNA) and Tail dependence network approach (TNA), at different observation scales are used here to quantify linear and non-linear correlation effect.

By employing data from 2014 to 2021 for two datasets (SSE180 and SSE 50), we compare the modified Herfindahl index, turnover, break-even transaction cost (BETC), drawdown and return characteristics of three asset allocation strategies. We show that incorporating both the idiosyncratic information of asset and systemic interconnectedness information derived from the network topology can substantially improve the return performances compared to two benchmark models. The significant performance improvement is also supported by consistent less drawdown. Clearly, utilizing the multi-objective portfolio optimization with idiosyncratic and systemic risks enables investors to obtain different portfolio configurations, and according to adjusting the parameter $\alpha$ based on the current market environment to achieve an optimal portfolio.

The results obtained in our study encourage the develop of further research line, on extending the approach by considering stock markets as directed networks, describing risks by various measures, for instance VaR ( See Pang et al. (2005)[41]), and quantifying the dependence structure of asset returns by diverse dependency measures, for example the mutual information to detect non-linear dependencies (See Fiedor et al. (2014)[42]). This framework may endeavor to enhance our understanding on risk transmission mechanism of the financial market and be a valuable method for portfolio selection.

**Acknowledgments**

This work is supported by Scientific and technological activities support funding for returned overseas students of Shaanxi province No. 27 and National Natural Science Foundation of China 71901171 and 72071006, Startup funding of NWPU G2021KY05101 and the interdisciplinary research fund for liberal arts research of NWPU 21GH031109.